\documentclass[twocolumn]{article}
\usepackage{graphicx}
\begin{document}

\title{A correlation between Earth's inclinations and the times of the cold events recorded in Devils Hole climate data}

\author{W. W\"olfli
\footnote{Institute for Particle Physics, ETHZ H\"onggerberg, CH-8093 Z\"urich, Switzerland. woelfli@phys.ethz.ch}
 \ and W. Baltensperger\footnote{Centro Brasileiro de Pesquisas F\'\i sicas, r.~Dr.~Xavier Sigaud, 150, Urca, 22290 Rio de Janeiro, Brazil,\qquad \qquad \qquad baltens@cbpf.br}}
\date{}
\maketitle

\noindent
{\bf The calculated values of Earth's inclinations during the cold events in the 500 kyr climate record of Devils Hole show a correlation: they cluster in the regions around 1 and 2 degrees. The Devils Hole record has been chosen, since it was dated by absolute methods. Other climate records co\-ver\-ing the same period also have a reduced number of cold events between the two regions. The correlation lends support to the proposal of R. A. Muller  and G.J. Mac\-Do\-nald  that the observed 100 kyr climate cycle is due to the varying inclination of Earth's orbit and to material located near the invariant plane which shields off the solar radiation.}

The climate of the late Pleistocene (ca. 800 - 11.5 kyr BP) shows a 100 kyr glacial cycle. Muller and McDonald \cite{Muller1} proposed that this feature is caused by variations of the inclination, i.e.~the tilt of Earth's orbital plane, rather than by eccentricity variations, as assumed in the Milankovitch theory. Their model implies that some gas or dust was located near the invariant plane, which reduced the solar radiation on Earth whenever the inclination was smaller than the angle delimiting this cloud.

\begin{figure*}
\includegraphics[width=16cm,keepaspectratio]{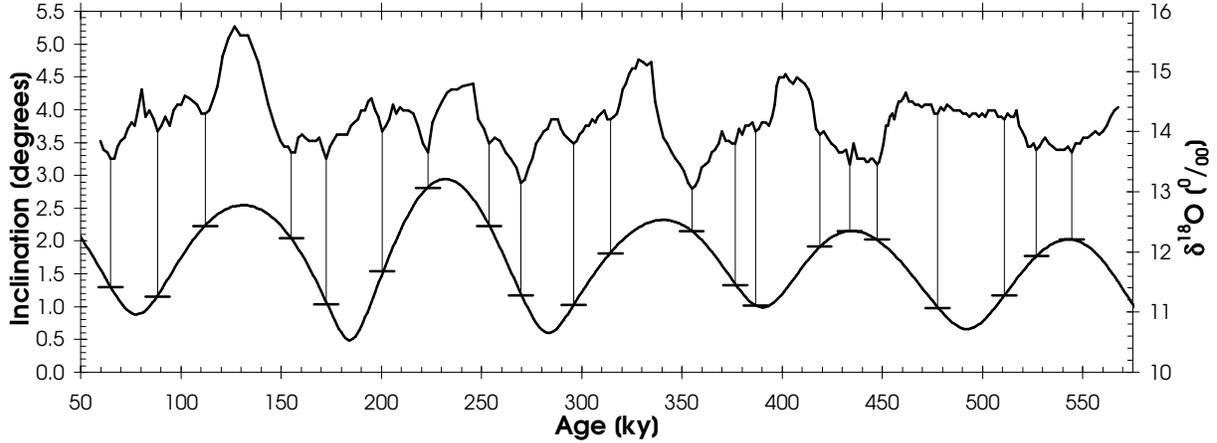}
\caption{$\delta^{18}$O climate data from Devils Hole \cite{Winograd} (upper curve) and Earth's orbital inclination (lower curve) over 600 kyr before present \cite{Muller1}. The vertical lines connect cold events with the corresponding inclination values.}
\end{figure*}
The present paper reports a correlation between the values of Earth's orbital inclination and the times of the cold events using the climate record of Devils Hole \cite{Winograd}. This record (Fig.~1) is chosen because the ages of the calcite specimens were determined by the U-Th method. The inclination of the Earth, also displayed in Fig.~1, was calculated by Quinn et al \cite{Quinn} and transformed into the invariable plane by Muller \cite{Muller2}. Fig.~2 shows that the inclinations at the times of the temperature minima mostly lie within two regions, which are centred at about 1 and 2 degrees, respectively. Fig.~2 indicates that the correlation does not depend on the age of the specimens. In most cases the inclinations have relatively rapid time dependences at the points of occurrence, so that these values are not favoured by long exposures. Admittedly, in some cases the selection of cold periods is arbitrary, however, with slightly different choices of temperature minima the correlation still persists.

Similar plots have been made with climate records, which were dated either by less reliable absolute methods or tuned to the Milankovitch set of parameters. In the Vostok \cite{Petit} data a reduced occurrence of inclinations between 1.4 to 2.0 degrees can be discerned. The records of Tiedemann et al. \cite{Tiedemann} and Specmap \cite{Imbrie} show the correlation during the first 0.6 Myr.  The GRIP \cite{GRIP} record over 150 kyr is more detailed, but too short in time for this kind of analysis. The many Dansgaard-Oeschger events are too fast to be explained by changes in Earth's inclination.

\begin{figure}
\includegraphics[width=7.7cm,keepaspectratio]{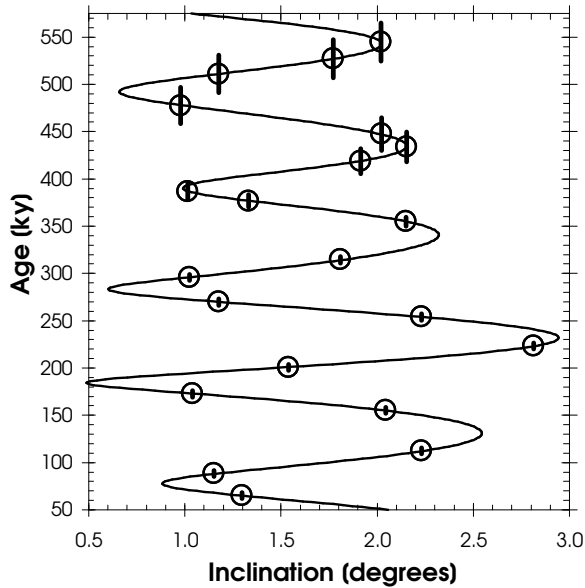}
\caption{Age of cold events versus inclinations. The former tend to cluster near 1 and 2 degrees. The vertical bars indicate the errors in the age, which, near 400 kyr,  increase from $\pm 3$ to $\pm 20$ kyr. Thus the significance of the oldest samples is small. The solid line represents the inclination.} 
\end{figure}

The correlation displayed in Fig. 2 indicates that Earth's inclination, which is not considered as a parameter of Milankovitch forcing, must be included into the discussion of the Ice Age problem. The gas is concentrated close to the surface of rotationally symmetric cones, which form angles of 1 and 2 degrees with the invariant plane, respectively. A distribution of dust, sharply peaked at angles of 2 and 10 degrees to the ecliptic, has been observed with the satellite IRAS \cite{Dermott} . The authors suggest that a cone could become a densely populated region of space, if there are orbiting particles, for which an observed angle is an extreme value of their time varying inclination. If during the Pleistocene a cloud reduced the insolation on Earth, it must have had a much higher density and a lifetime short compared to 10 kyr. Candidates for the constituents of such a cloud are atoms or ions, whose first excitation lies outside the main spectrum of the solar radiation. The conical distribution may result from their properties, in particular the ratio between the force from the solar radiation and the gravitational attraction.  A possible source of the gas is described in ref.\cite{Woelfli}. 

\section*{Acknowledgement}
The authors thank H.-U. Nissen for helpful discussions.

\end{document}